# Giant Rydberg Excitons in Cuprous Oxide

T. Kazimierczuk[1], D. Fröhlich[1,*], S. Scheel[2], H. Stolz[2], M. Bayer[1]

[1]Experimentelle Physik 2, Technische Universität Dortmund, D-44221 Dortmund, Germany.

[2]Institut für Physik, Universität Rostock, D-18051 Rostock, Germany.

*Correspondence to: Dietmar.Froehlich@tu-dortmund.de

**Highly excited atoms with an electron moved into a level with large principal quantum number are fascinating hydrogen-like objects. The giant extension of these Rydberg atoms leads to huge interaction effects. Monitoring these interactions has provided novel insights into molecular and condensed matter physics problems on a single quantum level. Excitons, the fundamental optical excitations in semiconductors consisting of a negatively charged electron and a positively charged hole, are the condensed matter analogues of hydrogen. Highly excited excitons with extensions similar to Rydberg atoms are attractive because they may be placed and moved in a crystal with high precision using microscopic potential landscapes. Their interaction may allow formation of ordered exciton phases or sensing of elementary excitations in the surrounding, also on a quantum level. Here we demonstrate the existence of Rydberg excitons in cuprous oxide, $Cu_2O$, with principal quantum numbers as large as $n=25$. These states have giant wave function extensions of more than 2 micrometers, compared to about a nanometer for the ground state. The strong dipole-dipole interaction is evidenced by a blockade effect, where the presence of an exciton prevents excitation of a further exciton in its vicinity.**

Problems in condensed matter physics inherently carry many-body character. To achieve transparent explanations, they are often mapped onto effective models involving elementary excitations to which particle character can be assigned. A particularly prominent example is the exciton [1] which has turned out to be an extremely successful concept in describing the optical properties of semiconductors and insulators. Within this concept, an electron is isolated in the conduction band containing a low carrier concentration. Simultaneously, a missing negative charge in the almost fully occupied valence band is treated as positively charged `hole`. The Coulomb attraction between electron and hole leads to formation of bound exciton states. The influence of the many-body system is comprised in quasi-particle properties such as the effective masses and in environmental parameters such as the dielectric constant.

The exciton obviously has strong similarities with the hydrogen atom, which has been substantiated by the observation of exciton states with binding energies $-\mathrm{Ry}/n^2$ below the band gap. Here Ry is the Rydberg energy and *n* is an integer, in analogy to the principal quantum number *n* of hydrogen. The observation of highly excited excitons is typically prevented by a small Rydberg energy of a few meV (e.g., 4.2meV in the prototypical semiconductor GaAs with *n*=3 as highest observed state), about three orders of magnitude smaller than in hydrogen. Therefore, these states are energetically spaced too closely to each other and to the ionization continuum to be resolvable.

For our search of giant excitons we have chosen the semiconductor cuprous oxide (Cu$_2$O), in which excitons were first observed [2,3], facilitated by the comparatively large Rydberg energy of around 100meV. Cu$_2$O has a direct band gap, see SI for details. The highest valence and the lowest conduction bands are formed from Cu-states, namely the 3d and 4s orbitals, respectively. Therefore, photon absorption leads to electron-hole pair creation at the same atom, different from the vast majority of other direct-gap semiconductors like GaAs for which conduction and valence band originate from the levels of different atoms.

The excitons associated with these two bands form the so-called yellow series because of their energies around 2.1eV, corresponding to 590 nm wavelength for excitation by light. Both bands have the same parity, and therefore electric dipole transitions, by which excitons with S-type envelope wave function would be created, are forbidden [4]. On the contrary, excitons with a P-envelope are dipole allowed as outlined in the Supplementary Information (SI). Already in early works by Gross et al., the P-exciton series could be followed up to $n = 9$ [2,3] and over the years could be extended to $n = 12$ [5]. Going beyond these $n$-numbers would allow one to create the solid state analogue to Rydberg atoms.

Rydberg atoms with sizes in the order of 100 nm have been intensively studied recently because of attractive properties such as long excited state life times, strong dipolar interactions etc., which might pave the way for quantum information technologies [6]. Very recently [7], coupling of the electron in a single Rydberg atom ($n = 202$, radius of 2μm) to a Bose-Einstein condensate was reported, through which a collective oscillation of the whole condensate could be triggered.

Excitons with properties similar to Rydberg atoms might open appealing perspectives, because their positions may be precisely controlled by microscopic potential landscapes created by electrical gates, for example. Thereby their interaction could be controlled on a detailed level. In our experiment the exciton spectrum in $Cu_2O$ is studied by high-resolution spectroscopy in which the photon energy of a laser with 5 neV linewidth (corresponding to about 1.2 MHz) is scanned across the energy range of interest and the transmitted laser intensity is measured with high sensitivity (see SI). Usually, the success of semiconductors is based on the extremely high crystal quality achieved by artificial fabrication. Amazingly, $Cu_2O$ artificial crystals are clearly inferior in quality compared to natural crystals. In our case a $Cu_2O$ crystal with a thickness of 34µm is cut and polished out from a rock taken at the Tsumeb mine in Namibia. The sample is held strain-free at a temperature of 1.2 K in superfluid Helium (see Fig. 1b,c and SI).

In the top panel of Fig. 1a we present the absorption spectrum of P-excitons obtained from the transmission experiments, revealing a large number of lines. To take a closer look at the high energy part, one has to zoom into the spectrum with increasing resolution as done in the lower panels. The exciton lines are labeled by the corresponding principal quantum numbers. Surprisingly, we can uniquely identify states with $n$ up to 25, far higher than ever reported for any solid state system [5]. The high $n$ immediately triggers the question about the extension of the exciton wave function. Using the hydrogen relation, the average radius $\langle r_n \rangle$ of an orbital with principal quantum number $n$ and angular momentum $l$ is given by $\langle r_n \rangle = \frac{1}{2} a_B (3n^2 - l(l+1))$, where $a_B$ is the Bohr radius and $l = 1$ for P-states [8]. The Bohr radius for P-excitons is $a_B = 1.11$nm [9]. For $n = 25$ we thus get $\langle r_{25} \rangle = 1.04$µm, corresponding to a huge exciton extension of more than 2µm. The exciton wave function therefore spans more than 10 billion crystal unit cells (each containing 4 copper and 2 oxygen atoms) and has an extension of about 10-times the light wavelength (see Fig. 1d).

The measured absorption lines exhibit an asymmetry with a steeper slope on the high energy flank. The asymmetry is due to interference of a discrete excitonic state and a continuum of states from interaction with optical phonons [10-12]. Each absorption line is well described by an asymmetric profile (see SI). From corresponding fits, the exciton resonance energies $E_n$ can be accurately determined. These energies are shown in Fig. 2a as function of $n$ in a double logarithmic plot. The resonance energies follow the Rydberg formula $E_n = E_g - \frac{Ry}{n^2}$ to a very good approximation, as shown by the fit in Fig. 2a. From the fit we obtain the bandgap energy[1] $E_g$ = 2.17208 eV and the Rydberg energy Ry = 92 meV. However, our high-resolution spectroscopy revealed a slight deviation from the exact hydrogen Rydberg series which can be incorporated by employing the concept of quantum defects (see SI) such that $E_n = E_g - \frac{Ry}{(n-\delta_l)^2}$ with $\delta_P = 0.23$.

The line widths $\Gamma_n$, shown in Fig. 2b, decrease with increasing $n$ down to a few µeV. For principal quantum numbers below about $n=10$, the line shapes can be well described by Lorentzians, suggesting a homogeneous broadening. Here, the data are in accordance with an inverse cubic law of $n$. For higher $n$, the lineshape changes and becomes more and more Gaussian, see Fig. 1a. This indicates that the homogeneous broadening is superimposed by crystal inhomogeneities that are monitored by the extended exciton wave functions. Still, the linewidth decreases with principal quantum number so that for the highest $n=24$ it is found to be as small as 3 µeV. The homogeneous linewidth $\Gamma_n$ can be used to derive estimates for the exciton lifetime $\tau_n$ in state $n$ using the relation $\tau_n \approx \hbar/\Gamma_n$ which for the inhomogeneous case gives lower limits for $\tau_n$. For the highest principal quantum numbers we obtain lifetimes on the order of nanoseconds from the µeV linewidths.

At first sight, these lifetimes appear surprisingly long because the huge wavefunction extension may cause the exciton to be fragile as it is confronted with multiple scattering possibilities in the crystal. For low excitation powers, carrier-carrier scattering can be neglected. However, two important pathways remain for the decay of the P states, apart from direct radiative recombination. One is relaxation into lower lying states by spontaneous emission of far infrared photons with a few tens of meV energy. As the spontaneous transition rate is proportional to the third power of photon energy, it is strongly suppressed for such transitions compared to the visible and can be as well neglected due to the expected microsecond decay times. The other pathway is relaxation by emission of optical phonons [10]. From the overlap between the initial and final state exciton wave functions, the relaxation rate for this scattering is expected to scale as $1/n^3$ [13], in accordance with the experimental findings for the homogeneous contribution.

---

[1] Since the experimental data are taken at a photon momentum $\hbar K$, the band gap energy at K=0 is lower from the given value by the kinetic energy $\Delta E = \hbar^2 K^2/2M = 24 \mu eV$, where the exciton mass M=1.56m$_0$.

From the giant extension, huge Coulomb interaction effects are expected which we access by studying the transmission as function of laser excitation power and therefore exciton density. The area of each absorption peak corresponds to the absorption strength and is determined by the exciton oscillator strength. The peak areas are shown in Fig. 2c as function of $n$ at an applied laser excitation power of $P_L$=20µW, corresponding to an intensity of 6µW/mm$^2$. We find that the peak area also scales as $1/n^3$, but only in the range up to $n=17$. This dependence confirms the theoretical analysis for isolated P excitons [1,14] from which one expects for the exciton oscillator strength a behavior proportional to $\frac{n^2-1}{n^5} \sim \frac{1}{n^3}$ for large $n$. However, there are pronounced deviations for $n>18$ at the used excitation power. There the peak areas are reduced by almost an order of magnitude compared to the expected values for the chosen excitation power.

To explore the origin of this reduction in absorption strength in more detail, we measured the peak area as function of excitation intensity. Fig. 3a shows corresponding absorption spectra from $n=12$ upwards. With increasing power, the absorption lines continuously decrease with the higher lying ones fading away first. The peak areas are plotted as function of excitation intensity in Fig. 3b, showing a drop of the area starting from a characteristic power level for each principal quantum number. The powers, at which the drop starts, shift to lower excitation intensity with increasing $n$. For the lowest shown exciton with $n=12$, the area decreases only for the highest applied excitation intensity. These results suggest that interaction effects between excitons are responsible for the reduction of absorption and increase of transmission. For larger exciton sizes the interaction effects set in at smaller exciton densities.

To explain the reduction, we propose a dipole blockade effect similar to the one observed for Rydberg atoms. The blockade arises from the dipole-dipole interactions between Rydberg excitons, depending strongly on their separation. If an exciton is created, the energy for exciting another exciton nearby is lowered by the dipole interaction energy, away from the narrow undisturbed absorption line. Thereby a dipole blockade is established: Resonant absorption and exciton creation are no longer possible in the blockade volume $V_{\text{blockade}}$ in which the dipole interaction energy is larger than the narrow absorption line width $\Gamma_n$.

As a consequence the absorption $\alpha$ at a given exciton density $\rho_X$ in the illuminated crystal volume is reduced by a factor $(1-\rho_X V_{\text{blockade}})$ compared to the absorption of the unexcited crystal $\alpha_0(\hbar\omega)$, which leads to $\alpha(P_L,\hbar\omega) = \alpha_0(\hbar\omega)(1-\rho_X V_{\text{blockade}})$. The exciton density $\rho_X$, on the other hand, is determined by this absorption $\alpha$ times the laser power $P_L$ deposited within the exciton lifetime $\tau_n \propto 1/\Gamma_n$ in the crystal: $\rho_X \propto P_L \alpha / \Gamma_n$. Inserting this relation for the exciton density and solving for the absorption $\alpha$ allows one to derive the following scaling law for the absorption as function of laser power:

$$\alpha(P_L, \hbar\omega) = \frac{\alpha_0(\hbar\omega)}{1 + S_n \cdot P_L}, \qquad \text{(Eq. 1)}$$

where the $S_n$ describe the efficiency by which the absorption at the energy of the exciton state $n$ is blocked through the presence of excitons in the very same state. Eq. (1) describes well the observed dependencies in Fig. 3b: By fitting the experimental data for the peak area, we can extract the $S_n$ which are shown for $n$ from 12 up to 24 by the symbols in Fig. 3c. In this high-$n$ exciton range, $S_n$ varies enormously with the principal quantum number increasing by more than three orders of magnitude. By fitting the data with a power function we find a dependence on the 10-th power of $n$.

For understanding the strong n-dependence, one has to consider possible dipole-dipole interaction mechanisms. At large separations they can be modeled by a van der Waals (vdW) form for the interaction energy $E_{vdW}(n) = -\frac{C_6(n)}{R^6}$, where $R$ is the distance between two P excitons in state $n$. For smaller distances the interaction becomes resonant and is better described by a Förster (F) type dependence $E_F(n) = -\frac{C_3(n)}{R^3}$. The dependences of the coefficients $C_6$ and $C_3$ on the principal quantum number $n$ can be obtained by second-order nondegenerate perturbation theory for the dipole-dipole interaction ($C_6$) and by a first-order degenerate perturbation calculation ($C_3$) for the Förster mechanism. One finds that $C_6$ varies with $n^{11}$, while $C_3$ scales as $n^4$ (for details see SI). The onset criterion for the blockade of the dipole interaction energy becoming larger than the absorption line width, $E_{F/vdW} > \Gamma_n$, gives us a critical blockade radius $R_c$ as $R_c(n) = \sqrt[q]{C_q(n)/\Gamma_n}$. From this radius the volume $V_{\text{blockade}} = \frac{4}{3}\pi R_c^3$ around the exciton that is blocked for absorption can be derived, see Fig. 3b.

According to the considerations above, the dependence of the blockade efficiency $S_n$ on the principal quantum number $n$ is determined by the product of blockade volume $V_{\text{blockade}}$ times the exciton lifetime $\tau_n \propto 1/\Gamma_n$. From $V_{\text{blockade}} \propto R_c^3 = \left(\sqrt[q]{C_q/\Gamma_n}\right)^3 \propto n^7$ for both possible mechanisms of dipolar interaction and $\Gamma_n \propto n^{-3}$ we obtain an extremely steep increase of the blockade efficiency $S_n$ with increasing principal quantum number $n$, given by its 10$^{\text{th}}$ power: $S_n \propto \dfrac{V_{\text{blockade}}}{\Gamma_n} \propto n^{10}$, in excellent agreement with experiment.

From these considerations we expect the Coulomb blockade to occur not only for excitons with the same $n$, but also for different $n$. In effect, the experiment so far resembled a single beam pump-probe experiment with degenerate pump and probe. To test our suggestion further, we implemented another tunable laser with a line width of 1 neV (about 250 kHz in frequency) into the setup, to be able to vary the pump and probe photon energies independently. In a first run, we kept the photon energy of the pump laser fixed at the position of the $n = 14$ exciton, while simultaneously sampling the exciton spectrum to higher $n$ with the probe laser (see Fig. 3d). Qualitatively similar observations as described above are made. With increasing pump intensity the transmission at all exciton lines - not only the resonant ones - increases, accompanied by some line broadening. This demonstrates that the blockade works also for off-resonant excitons[2]. Furthermore, at a fixed excitation intensity excitons with high principal quantum numbers are more strongly blocked by the off-resonant pump than lower lying ones. Within the excitation spot the number of excitons and the separation between them varies, contributing to a broad absorption background in the 2-color studies due to the widely varying exciton interaction energies. The remaining absorption line arises from excitons with interaction energies below the linewidth.

---

[2] By monitoring the crystal transmission under modulated-in-time laser excitation, we verified that the excitation induced changes of the absorption do not arise from long-lived exciton populations formed after relaxation of optically injected excitons. An example for such a long-lived exciton is the paraexciton which is the spin-triplet configuration of the 1S ground state exciton with lifetimes in the µs-range.

In a next step the pump was scanned, and the probe photon energy was kept at the energy of the *n*=17 exciton, for which we monitored the change of absorption. The goal was to detect a reduced absorption and hence increased transmission whenever the pump photon energy hits an exciton resonance. To demonstrate the effectiveness of the blockade, the pump photon energy was varied from *n*=6 to *n*=10 with Bohr radii up to about 100 nm. Whenever the pump laser hit a narrow exciton resonance in this energy range the transmission of the probe laser increased, as demonstrated in Fig. 3e. This proves that excitons in lower principal quantum number states stay intact and prevent exciton creation in the *n*=17 state.

Since the efficiency by which an exciton blocks exciton creation in its vicinity is determined by its extension, we corroborate our interpretation by deliberate variation of the exciton size. An efficient tool for such a size variation is the application of a magnetic field, in our case along the optical axis (taken as z-axis). The magnetic field leads to a splitting of exciton levels with different magnetic quantum numbers by the Zeeman effect. Due to the resulting complexity we do not discuss the splitting pattern here in detail. Instead we focus on the Rydberg blockade related wave function engineering. In that respect most relevant is that the magnetic field adds parabolic confinement potentials for electron and hole to the exciton Hamiltonian $\frac{e^2 B^2}{8m}(x^2 + y^2)$, whose strength can be controlled by the field strength. As a result the exciton wave function is squeezed. The characteristic length scale for the magnetic confinement is the magnetic length $\ell_c = \left(\frac{\hbar}{eB}\right)^{0.5}$, which for 1 T field strength is 25.65 nm underlining the strong impact even for excitons with intermediate principal quantum numbers.

Figure 3f shows the absorption at the position of the $n=15$ exciton, recorded at $B=0$ and $B = 0.8$T as function of the applied optical excitation intensity. Note, that the shift of the $n=15$ exciton in magnetic field was carefully monitored, so that the excitation laser could be tuned into exact resonance with the exciton as in the zero field case. While without magnetic field these states become strongly bleached and can hardly be resolved for laser powers exceeding 100 µW, in magnetic field the drop of absorption with increasing laser power is much weaker, so that even at the highest applied laser intensity of 20mW/mm$^2$ the absorption is about half of the low intensity value. This finding is also in full accord with the expectation for the Rydberg blockade mechanism.

We are convinced that the observation of these giant excitons opens a new field in condensed matter spectroscopy. For example, the blockade may be exploited in applications such as nonlocal all-optical switching or mesoscopic single-photon devices. Furthermore, it will be interesting to work out similarities and differences of Rydberg atoms and Rydberg excitons. Important differences may arise from the considerably lower exciton Rydberg energy, potentially reducing the thermal stability. On the other hand, the resulting exciton Bohr radius is much larger for similar principal quantum numbers, so that similar blockade volumes may be reached for excitons at considerably smaller $n$ than for atoms. In parallel, the light wavelength is shrunk in the Cu$_2$O crystal by the refractive index of 3, so that Rydberg excitons may allow testing the light-matter interaction description such as the electric-dipole approximation. A further difference is the very different lifetime of the Rydberg excitons which are in the ns-range for the excitons while they may reach ms for atoms, which causes also the linewidths to differ significantly. This lifetime is influenced by the radiative decay of the exciton due to the much larger oscillator strength and is also influenced by phonon relaxation absent in the atomic case.

These differences should be reflected also in studies of interaction effects among excitons which may be deliberately induced and controlled by exciting additional excitons in particular $n$ states. The relaxation of excitons by phonons may lead to the formation of low-$n$ exciton populations with which the Rydberg excitons may interact. One of these excitons, the paraexciton is considered to be a prime candidate for Bose-Einstein condensation in cuprous oxide [15], so that recent experiments in atomic physics [7] could be mimicked by studying the interaction of a Rydberg exciton with such a condensate.

Also problems of molecular physics might be addressed as for Rydberg atoms: The possibility to excite molecules from two excitons in $Cu_2O$ has been discussed for a long time [16] but so far such molecules have not been demonstrated. Rydberg excitons open new perspectives here due to their strong dipole-dipole interactions [17]. Exciton molecules with varying constituents could be excited. Tuning of $n$ may allow tuning of the binding energy of exciton molecules [18-21]. Also the number of excitons forming a molecule may be varied up to large cluster-like states or extended condensed phases. Due to the crystal environment, Rydberg excitons could allow studies hardly possible in atomic or molecular physics. For example, the position of individual Rydberg excitons might be accurately controlled by applying spatially modulated strain fields to the crystal. Bringing the studies down to the level of a single exciton with controlled position will allow quantitative assessment of the interaction energies. Furthermore, additional electric or magnetic fields may be applied by which the interaction between Rydberg excitons as well as their stability could be controlled dynamically.

Subjecting Rydberg atoms to high magnetic fields allowed mimicking hydrogen atoms [22] in white dwarf stars which represent a non-integrable problem leading to chaotic behaviour. Due to the small Rydberg energy this regime should be possible to enter for Rydberg excitons already in the low field regime achievable with standard permanent magnets.

**Contributions:**

T.K., D.F. and M.B. conceived, designed and carried out the experiments mentioned in this manuscript. H.S. and S.S. contributed through the model for the Rydberg blockade. All authors cooperated in data analysis, discussions and preparation of the manuscript.

**Competing financial interests:**

The authors declare no competing financial interests.

**Acknowledgments:**

We thank R. Hönig for experimental support with the first measurements. We acknowledge the financial support by the Deutsche Forschungsgemeinschaft (BA 1549/18-1 and SFB 652 "Strong correlations and collective effects in radiation fields").

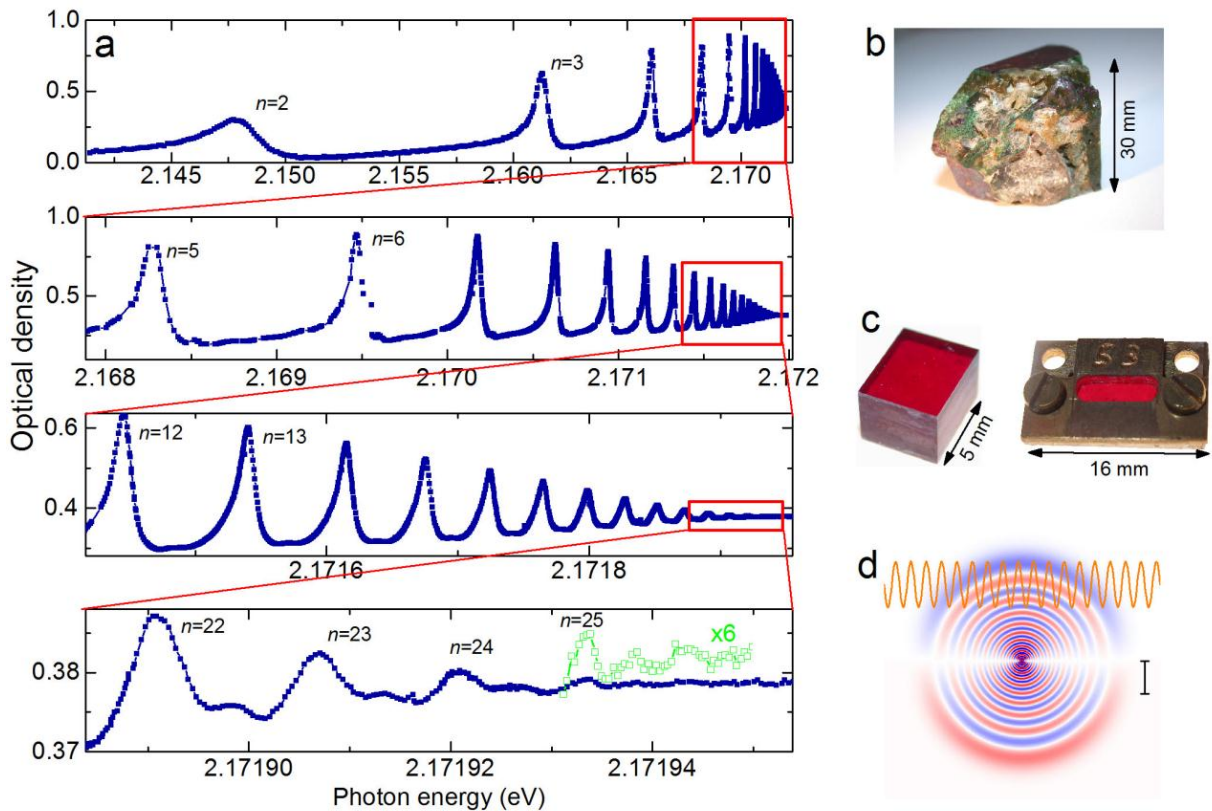

**Figure 1. High resolution absorption spectra of yellow P excitons in $Cu_2O$.** (a) Spectra are measured with a single frequency laser on a natural sample of 34μm thickness at 1.2K. Peaks correspond to resonances with different principal quantum number $n$. Consecutive panels present close-ups of the corresponding areas marked by rectangles in the corresponding upper panel. On the right-hand side: (b) Photograph of natural $Cu_2O$ crystal, from which samples of different size and crystal orientation were cut. (c) Large crystal and thin crystal mounted strain-free in a brass holder. (d) Wavefunction of the 25P exciton. For visualization of the giant extension, the corresponding light wavelength is shown as the period of the sine function. The bar corresponds to the extension of 1000 lattice constants.

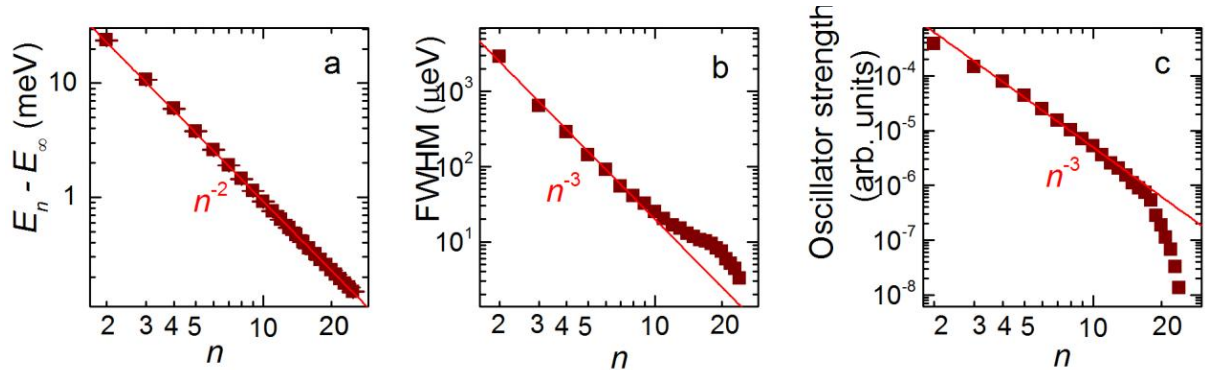

**Figure 2. Dependences of experimentally determined parameters of P exciton lines as function of principal quantum number *n*, evidencing power law behavior.** (a) Exciton binding energy: squares give resonance energies $E_n$, solid line represents the $n^{-2}$ dependence expected from the Rydberg formula with Ry=92meV, and $E_\infty = 2.17208\,\text{eV}$. (b) Squares are experimental values of absorption line width, solid line shows $n^{-3}$ dependence. The slight deviation of the experimental data from this dependence to higher values for *n* exceeding 10 may be due to some influence of crystal inhomogeneities in the illuminated volume. (c) Squares give experimental data of oscillator strength (peak area) in arbitrary units, solid line shows $n^{-3}$ dependence expected for single non-interacting exciton.

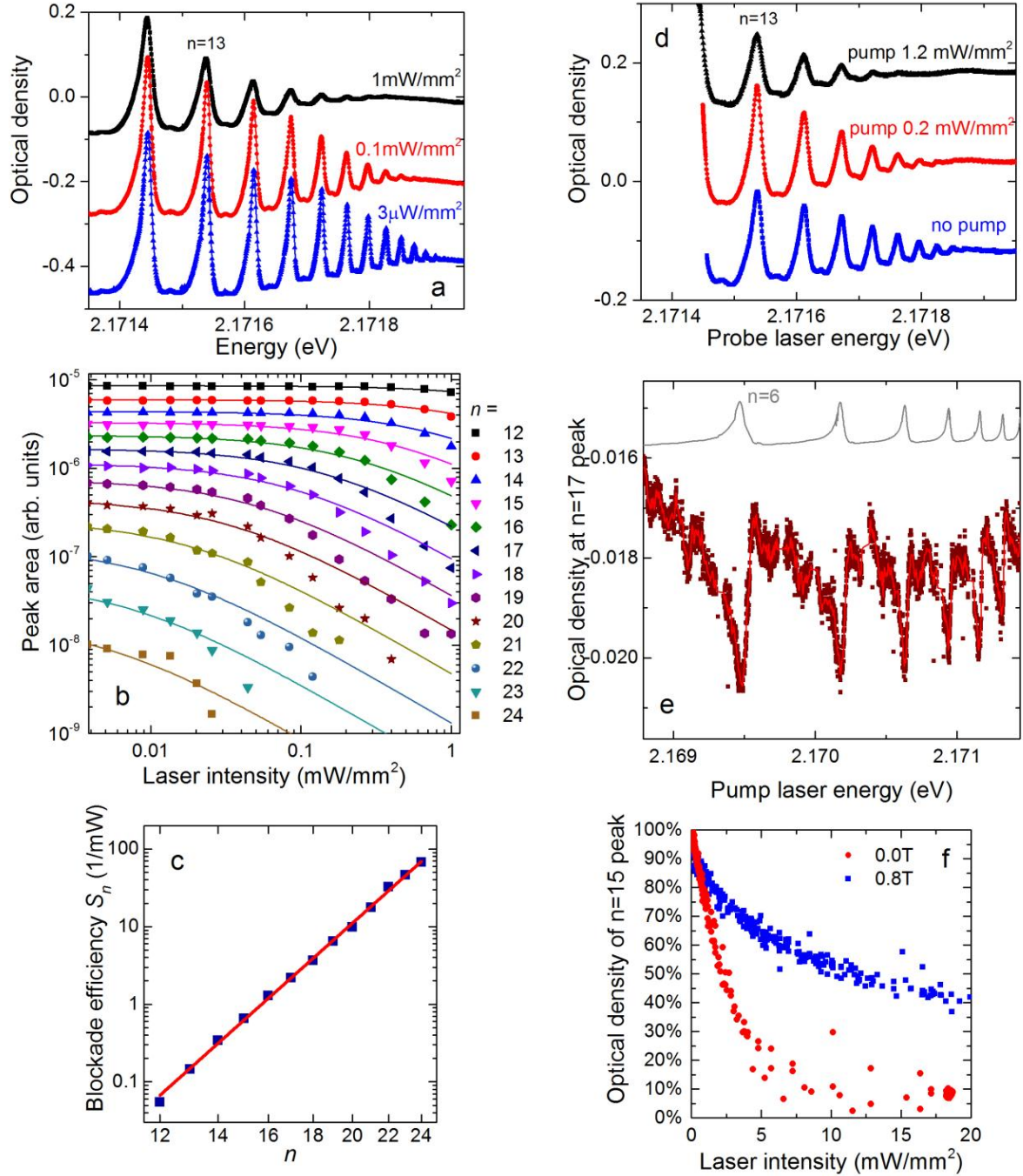

**Figure 3. Reduction of excitonic absorption due to dipole blockade.** (a) Absorption spectra measured with different laser intensities. Note the quenching of the high *n* resonances when applying stronger laser excitation. (b) Dependence of oscillator strength (peak area) on laser power for different *n* resonances, solid lines show $\frac{A}{1+S \cdot P_L}$ fits to the data. (c) Dependence of blockade efficiency $S_n$ on laser power, solid line shows fit according to $n^{10}$ dependence. (d) Absorption spectra in the two beam experiment. The presented curves were measured with different powers of the pump laser and constant probe laser power. The energy of the pump laser was fixed at $n=14$. (e) Change of absorption at $n=17$ resonance in the two beam experiment as function of the pump laser photon energy (pump intensity 0.3 mW/mm$^2$). The red line is a guide for the eye. The grey line presents a single-beam absorption spectrum in this energy range, showing that the absorption is quenched stronger when the pump laser is tuned to an exciton resonance. Note the slight line shifts between the resonances in the two spectra arising from exciton-exciton interaction. Due to the variation of the exciton separation in the laser spot, these shifts cannot be assessed quantitatively, for which one would have to control the exciton position. (f) Influence of the magnetic field on the absorption reduction effect. Shown are the optical densities at $B=0$ and $B=0.8$T as function of excitation power.

# Supplementary Information for "Giant Rydberg Excitons in Cuprous Oxide"

First we present details of the material Cu$_2$O and its electronic structure with emphasis on the lowest exciton series (yellow series). We then describe the experimental setup and give the corrections of the nP resonances due to phonon background which allows us to table the P exciton energies. In the next section we discuss details of the theoretical estimate of the dipole blockade efficiency and the blockade volume.

**Cuprous oxide**

Cu$_2$O crystalizes in a cubic lattice (space group O$_h$). The crystal unit cell and the electronic band structure are shown in Fig. S1. The two upmost valence bands ($\Gamma_7^+$ and $\Gamma_8^+$ symmetry) stem from Cu 3d electrons, the lowest conduction band from Cu 4s ($\Gamma_6^+$ symmetry) and the next higher conduction band from O 2p electrons ($\Gamma_8^-$ symmetry). Excitation from the two valence bands to the two conduction bands results in four exciton series, which are named yellow, green, blue and violet series, as shown in Fig. S1. We are concerned with excitons from the lowest energy series ($\Gamma_7^+$ to $\Gamma_6^+$ transition). The exciton motion can be divided into the centre of mass motion and the relative motion of electron and hole. The relative motion is characterized by the principal quantum number *n* and the orbital angular momentum quantum number *l*=0,1,2, … (labelled S,P,D,… as in hydrogen). Because of its band structure, Cu$_2$O has rather unique properties as compared to other direct-gap semiconductors like e.g. GaAs. One unusual property is that the upmost valence and the lowest conduction band are derived from states of the same ion (see above). Because both bands have the same parity, dipole transitions between band states are forbidden. Transitions from the ground state (full valence band, empty conduction band) to excitons, however, obtain the additional quantum numbers of the relative motion of electron and hole. The total symmetry of the exciton is thus given by the direct product of the symmetries of the bands and the envelope [23]:

$$\Gamma_{ex} = \Gamma_v \otimes \Gamma_c \otimes \Gamma_{env}$$

$$= \Gamma_7^+ \otimes \Gamma_6^+ \otimes \begin{cases} \Gamma_1^+ & \text{for S envelope} \\ \Gamma_4^- & \text{for P envelope} \\ \Gamma_3^+ + \Gamma_5^+ & \text{for D envelope} \\ \vdots & \end{cases}.$$

Because of parity, S excitons are dipole forbidden but quadrupole allowed, whereas P excitons are dipole allowed [24].

Because of their coupling to light, the excitons discussed so far (free excitons) in general have to be considered as polaritons. Polaritons are exciton – photon mixed states [25] that move through the crystal with their group velocity $v_g$, which can be derived from their dispersion relation. For the quadrupole allowed 1S excitons of the yellow series the polariton character was clearly demonstrated [26-28]. Although our results are well described within the exciton picture, it will be interesting to analyse the P series within a multi-resonant polariton concept [29].

**Experimental setup**

In Fig. S2 we present the schematic setup of our experiment. The setup serves two purposes:

i) classical transmission spectroscopy by use of a white light source and a monochromator,

ii) high resolution spectroscopy by use of single frequency dye lasers.

We started our experiments as reported earlier in literature [5,12] by use of a broadband white light source (Energetiq LDLS) in front of the sample and a double monochromator in second order with a CCD camera as a detector (spectral resolution of about 10μeV). In our first try we detected resonances only up to $n=12$. We discovered, that broadband white light leads to a quenching of higher excitons, probably due to the fact that, besides the simultaneous excitation of all P excitons from the yellow series, also excitons from the other series (green, blue and violet series, see Fig. S1) and even free carriers beyond the band gap are excited, which leads to multiple interactions, potentially destroying high-$n$ excitons. Inserting cut-off and interference filters in front of the sample brought some improvement (resonances up to $n=13$ were detected). After inserting a small double monochromator (Solar MSA-130, bandwidth about 0.1nm ~ 300 μeV), we were able to detect resonances up to $n=17$ (Fig. S3). Because of the limited spectral resolution of the detection monochromator (about 10 μeV) these resonances are broader than the resonances determined with the laser. The limitation of the bandwidth of the illumination monochromator to about 300μeV was a possible reason, why only resonances up to $n = 17$ were resolved.

We then decided to implement a single frequency dye laser. Most of the measurements were done using a Coherent 899-21 laser with a linewidth of about 5neV corresponding to 1.2 MHz. We first focused the laser down to about 50μm spot size. In that way we obtained resonances up to $n = 25$, if the laser power was cut down to about 500nW. A single frequency laser has a broad band background due to ASE (amplified spontaneous emission). Sending the laser through the Solar-monochromator did not bring any further improvement. We improved the uniformity of the laser cross-section by inserting a fibre and obtained the best results by use of a 2mm spot size of the laser on the crystal. This rather large spot size has the advantage, that laser power levels of several µW could be used without running into any detectable exciton blockade effects. For the balanced detection system (NewFocus Nirvana 2007) power levels of $> 5\mu W$ were necessary to guarantee linearity and low noise operation. The balanced detection led to a suppression of laser fluctuations to less than 0.1%. A careful alignment of sample and reference beam on the detector was important to achieve sufficient resolution of the small absorption signals. We tried several different beam splitter setups. We finally decided to use a Glan prism as a beam splitter. The laser system allows an automatic tuning of 20GHz (83µeV) only. The spectral range from 2P up to 25P, however, corresponds to 27meV. Besides the electronically controlled thick etalon there were two other rougher tuning elements (Lyot-filter and thin etalon). The tuning of the laser was automated with use of a stepper motor for the micrometer screw of the Lyot-filter and also for the potentiometer of the thin etalon. The wavelength setting was controlled by a high resolution wavemeter (HighFinesse WSU). All tuning and diagnostic elements were implemented in a LabView program, which allowed us to control spectral range, scan speed, etc. For the measurements of the intensity dependence (Fig. 3), it was of importance to stabilize the laser power. For that purpose we used a liquid crystal-based noise eater BEOC LPC.

    The bulk of experience gained in optimizing the single color transmission studies could be exploited in setting up the two-color pump-probe experiment. For these experiments we additionally used a second tunable single-frequency dye laser Matisse DS with a line width of 1 neV, corresponding to about 250 kHz band width. One of the two lasers implemented then in the setup was used to measure transmission through the sample as described above, while the other

was focused at the same spot but not detected after passing through the sample. The overlap of the two beams was controlled by use of a 50μm pin-hole.

The samples were strain-free mounted in an Oxford Variox cryostat. For the magnetic field measurement up to 7T we used an Oxford Spectromag 4000 cryostat. The cryostat was fitted with a split-coil superconductive magnet. In the reported measurements a magnetic field of 0.8T was applied in the Faraday configuration, i.e. the field is oriented parallel to the optical axis.

**Corrections of nP resonances due to phonon background**

The measured absorption lines were fitted using the formula [10-12]:

$$\alpha_n(E) = C_n \frac{\frac{\Gamma_n}{2} + 2q_n(E - E_n)}{\left(\frac{\Gamma_n}{2}\right)^2 + (E - E_n)^2}, \qquad \text{(Eq. S1)}$$

where the amplitude $C_n$ is proportional to the oscillator strength, the width $\Gamma_n$ is related to the exciton lifetime through $\tau_n = \pi\hbar/\Gamma_n$, the exciton resonance energy $E_n$ is taken without the phonon background, and the parameter $q_n$ describes the line asymmetry. In Table S1 we present the experimentally measured positions of the absorption lines and the exciton energies $E_n$ after unfolding using Eq. S1. For $n = 2$ the relative shift from the unfolding amounts to more than 1meV and for higher-$n$ excitons it strongly decreases similar to the linewidth (i.e., roughly proportional to $n^{-3}$).

**Description of exciton series by Rydberg formula including quantum defects**

As shown in the manuscript for the P-excitons, the exciton energies within a series of states with a particular angular momentum but varying *n* is very well described by the Rydberg formula $-\mathrm{Ry}/n^2$.

In hydrogen, levels of the same principal quantum number *n* are degenerate with respect to the orbital angular momentum *l* (neglecting the Lamb shift). In cuprous oxide, however, we find in combination with data recorded earlier [25] for a fixed *n* small, yet systematic splittings between states of different angular momentum which arise from a deviation of the electron-hole interaction potential from a pure 1/r-dependence due to the crystal environment.

The deviation arises from effects such as the central-cell correction due to band non-parabolicity, the electron-phonon interaction etc. Deviations from the Rydberg formula are known also from the physics of Rydberg atoms: The spectrum of hydrogen-like (alkali) atoms differs from the spectrum of hydrogen for low angular momentum states because the valence electron interacts with the ionic core. The Pauli repulsion and the dipole polarization of the rump ion's electron cloud leads to a larger binding energy compared to hydrogen.

From atomic physics it is established that the deviations of the energy levels from the Rydberg formula can be accounted for by the concept of quantum defect in which in the Rydberg formula the following replacement is made $n \rightarrow n-\delta_l$, where the $\delta_l$ is the quantum defect [8]. Also in the context of semiconductors the quantum defect concept has been already introduced theoretically [30]. Hence, the resonance energy is given by

$$E_{n,l} = E_g - \frac{\mathrm{Ry}}{(n-\delta_l)^2}. \qquad (\text{Eq. S2})$$

By fitting Eq. S2 to our data for the P excitons we find a quantum defect for the P series of $\delta_P = 0.23$ which describes the resonance energies of states with n>10 extremely well. The rudimentary data available for S excitons [25] provide a conservative estimate of $\delta_S \geq 0.35$.

**Estimation of dipole blockade radius**

A rough estimate of the exciton blockade volume can be obtained from the absorbed laser energy, given by the absorbed laser power times the exciton lifetime for which we use an upper boundary of 1 ns, based on the linewidth data. A pronounced reduction of the absorption is obtained already for 100 µW absorbed power, resulting in an absorbed energy of about 100 femtoJoule. With an exciton energy of more than 2 eV we obtain a number of about 300 000 excitons within the laser excitation spot, having a diameter of about 2mm, corresponding to one exciton in an volume of 300 µm$^3$. From this we derive a blockade radius of about 5µm.

This estimate of the dipole blockade volume can be done more quantitatively either based on the experimentally measured saturation under strong laser excitation or based on theoretical calculations of the dipole-dipole interaction. In case of the experiment-based estimate we use an expression for the blockade efficiency $S_n$ derived in the same way as Eq. 1 of the manuscript:

$$S_n = \frac{\eta \alpha_0 V_{blockade} T_1}{A_{exc} \hbar \omega},$$

where $\alpha_0 \approx 20$ mm$^{-1}$ is the peak absorption without the blockade, $V_{blockade}$ is the volume around the exciton unavailable for absorption, $A_{exc} \approx 3$ mm$^2$ is the laser spot area, $\hbar\omega \approx 2.1$ eV is the energy of the exciton and $T_1$ is the $n$-dependent exciton lifetime, assuming that the linewidth is radiatively limited. The lower bound for $T_1$ is given by the exciton linewidth: $T_1 \geq \frac{2\hbar}{\Gamma_n} \approx$ $7 \cdot 10^{-14}\text{s} \cdot n^3$, using the experimental data. The factor $\eta = 0.5$ accounts for the reflection losses on the cryostat windows and the crystal surface. For simplicity we assume here a constant value of $\alpha_0$. This simplification corresponds to the assumption that the deviation from the power law for highest-$n$ excitons (as shown in Fig. 2 in the manuscript) is due to local inhomogeneity of the crystal. By solving the above equation for $V_{\text{blockade}}$ we get:

$$V_{\text{blockade}} = \frac{S_n \cdot A_{exc} \hbar \omega}{\eta \alpha_0 T_1} \leq \frac{1.3 \cdot 10^6 \, \text{mW} \cdot \mu\text{m}^3}{n^3} S_n.$$

From inserting the fitted values for $S_n$ (see manuscript) into this formula we obtain the blockade volumes shown in Fig. S4 as function of the principal quantum number. The blocked volume increases drastically with $n$ by more than two orders of magnitude from $n=12$ upwards.

The second approach to estimate the blockade volume is based on an analysis of the dipole-dipole interaction between two excitons at a distance $R$. The dipole-dipole interaction may be discussed either in the resonant or nonresonant regime and can be computed either from perturbation theory or from an exact diagonalization of the static dipole-dipole interaction potential. The former method can be used to obtain the scaling of the interaction with the principal quantum number $n$: At large spatial separations between excitons, the dipole-dipole interaction contributes to nondegenerate second-order perturbation theory $\sim |<V_{\text{dd}}>|^2/\Delta$, where $\Delta$ is the energy difference between neighboring dipole-allowed states. The strongest contribution comes from coupling with the states of the same $n$, which would be degenerate for a perfect 1/r potential. As mentioned, lifting of the degeneracy between those states can be described using the quantum defect concept, according to which the resonance energy is given by Eq.S2.

The energy difference between neighboring dipole-coupled states scales as $n^{-3}$, both in case of the same $n$ and different $n$ states. The dipole moments increase as $n^2$, similar to the size of a Rydberg state. The dipole-dipole interaction $V_{dd}$ potential is quadratic in the dipole moments and scales with the third power of the distance between interacting excitons: $V_{dd} \propto \frac{p_1 p_2}{R^3}$, where $p_1$, $p_2$ are the dipole moments of the interacting excitons. Hence, the nonresonant van der Waals interaction takes the form:

$$E_{vdW}(n) \propto \frac{|V_{dd}|^2}{\Delta} \propto \frac{(n^4/R^3)^2}{n^{-3}} \propto \frac{n^{11}}{R^6}.$$

The proportionality constant $C_6(n=1)$ can be calculated by adding up the contributions from all dipole-coupled states.

On the other hand, for closer separation of the excitons, i.e. when the dipole-dipole interaction becomes comparable or larger than the energy spacing between the exciton levels, the interaction contains also resonant contributions proportional to $<V_{dd}>$. By the above arguments, this means that this Förster-type interaction takes the form:

$$E_F(n) \propto V_{dd} \propto n^4/R^3.$$

In this case the proportionality constant $C_3(n=1)$ can be calculated by diagonalizing dipole interaction within the subspace of dipole states considered as degenerate.

As discussed in the manuscript, we define a blockade radius by comparing the dipole-dipole interaction strength with the linewidth of the excitonic line, which scale falls into the Förster regime. The $C_3$ coefficient for the van der Waals interaction of two particular P states can be computed from the eigenvalues of the (static) dipole-dipole interaction potential in the two-exciton basis containing the nearest dipole-allowed states. In our calculation we include S and D states with the same and adjacent $n$. We neglect any retardation effects that might contribute at larger separations. Taking into account that the excitons reside in a host medium with a static permittivity of $\varepsilon = 7.5$ reduces the $C_3$ coefficient from its nominal free-space value. Thus, we estimate the $C_3$ coefficient for the resonant interaction between two excitons in Rydberg states as as $C_3/n^4 \approx 6 \cdot 10^{-4}$ µeV µm³. Consequently, the expected blockade volume is:

$$V_{\text{blockade}} = \tfrac{4}{3}\pi \frac{C_3}{\Gamma_{n/2}} \approx 3 \cdot 10^{-7} \; \mu m^3 \cdot n^7.$$

The results of these calculations are shown by the open symbols interpolated by a solid line in Fig. 3(d). Here we take into account the influence of a single exciton only onto the absorption. For the van der Waals form of the interaction the dependence looks similar. The calculations reproduce the increase of the blocked volume with increasing principal quantum number quite well. The calculated volumes are comparable to the experimentally estimated ones even though they tend to be somewhat smaller by a factor of 2 compared to the estimates from experiment. This is still a reasonable agreement, given the simplicity of our estimates done on purpose for transparency of the model, without trying to enforce agreement. For $n = 24$ the radius blockade volume would be about 10 µm, almost an order of magnitude larger than the exciton Bohr radius of this state, suggesting an enormous strength of the dipolar interactions.

**Exclusion of alternate explanations for the weak absorption of high-n excitons**

As discussed in the main text, the absorption strength for high-n excitons is weaker than the extrapolation of the dependence for low-n excitons, which follows the $n^{-3}$ scaling expected for isolated P excitons. We have attributed this to the effect of the dipole blockade. For completeness, we discuss and exclude possible alternative explanations.

At first sight, one may attribute the drop to thermal effects. The energy separation between the highest P excitons ($n = 25$) and the ionization continuum amounts to only about 140 µeV, which is comparable to the thermal energy at 1.2K. Yet their line width is below 3 µeV.

- 10 -

This shows that scattering of these Rydberg excitons by phonons and thermal excitation into the ionization continuum are not efficient. In any case thermal scattering would only broaden an absorption line, but not reduce its area. Therefore it can be safely ruled out as origin for the absorption drop.

The reduction may also point towards an invalidity of the electric dipole approximation, which is based on the assumption that the product of the light wave number $k$ and the extension of the state wave function $r$ is small compared to unity ($kr\ll 1$). Usage of this approximation may be challenged due to the large exciton size being comparable to or even considerably larger than the light wavelength. While this issue is worth further investigation, any influence related to it would not depend on the laser intensity, as observed in the experiment. The same argument applies to the temperature-based mechanism discussed above, because the laser powers causing bleaching of high-n excitons are well below the power levels needed to induce any measurable heating effects.

**Supplementary References:**

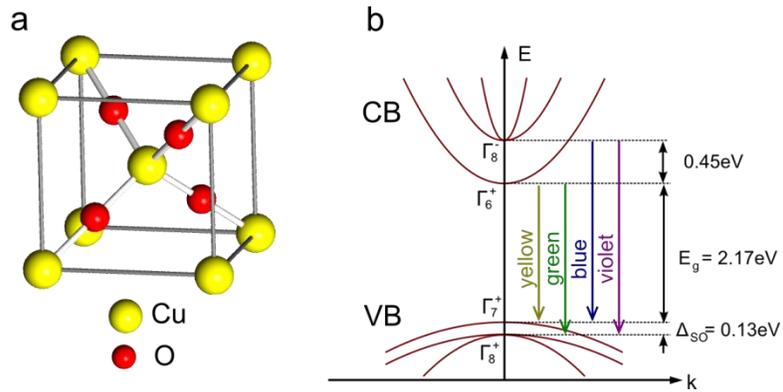

**Figure S1. Structure of Cu$_2$O.** (**a**) Elementary cell of Cu$_2$O: The O-atoms (red spheres) are arranged in a bcc lattice, the Cu-atoms (yellow spheres) form a fcc lattice. The sub-lattices are shifted relative to each other by one quarter of the body diagonal, which corresponds to cubic symmetry O$_h$. (**b**) Schematic electronic band structure around the center of Brillouin zone (Γ-point) in Cu$_2$O. Here the Γ$_i$ denote irreducible representations of the band functions for symmetry O$_h$ [23] not to be mixed with the line widths of the absorption lines. Transitions between the two valence bands (VB) and the two conduction bands (CB) lead to four exciton series denoted yellow, green, blue and violet.

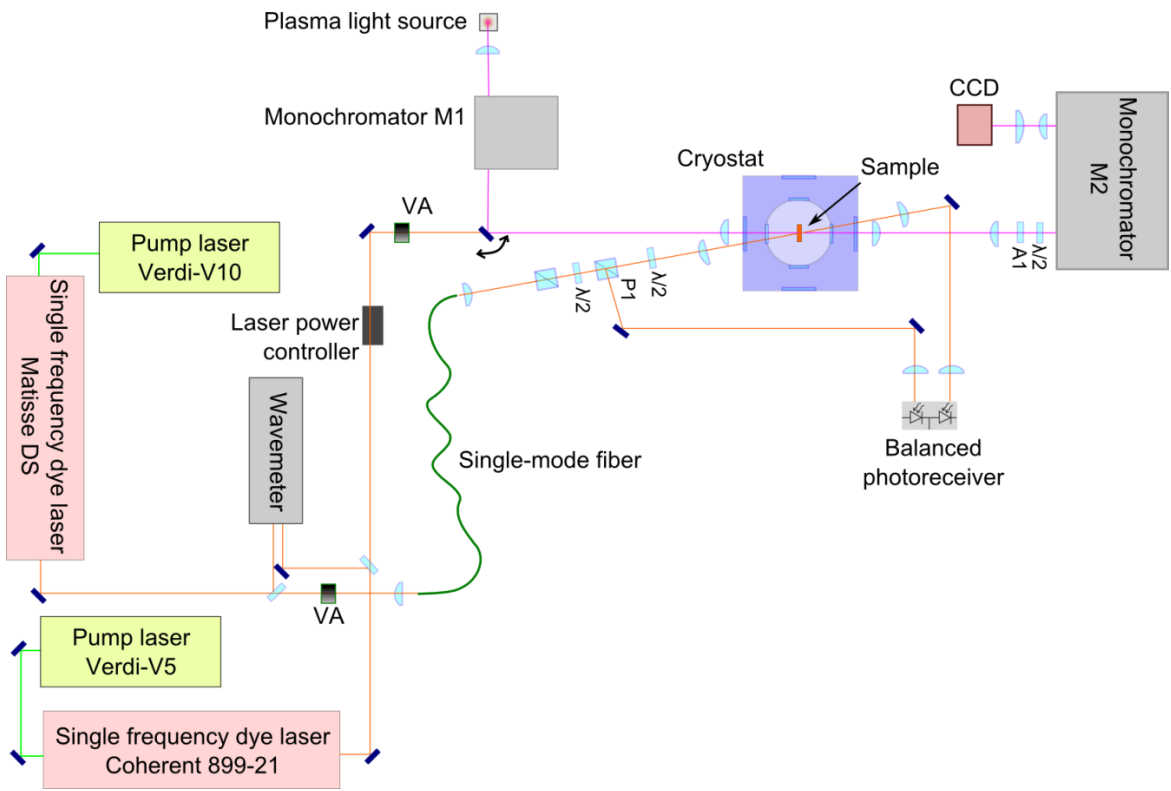

**Figure S2. Scheme of the experimental setup.** A1, film analyzer; CCD, charge-coupled device camera; P1-P2, Glan polarizer; λ/2, half wave plates; cryostat, Oxford Variox or Oxford Spectromag; VA, variable attenuators.

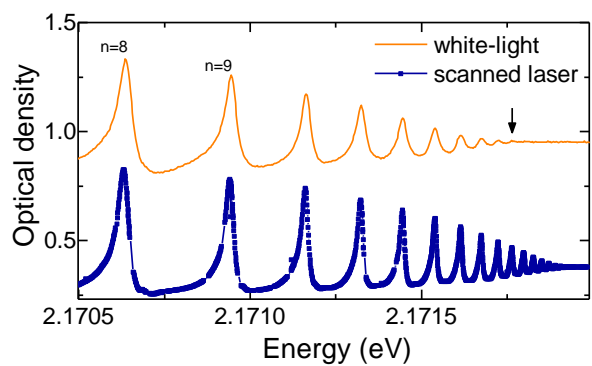

**Figure S3. Absorption spectra of P excitons of $Cu_2O$ from a 34μm thick sample at 1.2K.** Upper trace: Spectrum measured using a white light source with a resolution of 10μeV. The highest observed resonance $n = 17$ is marked with an arrow. Lower trace: Spectrum measured with a single frequency dye laser with a line width of 5neV, $n = 25$ is the highest resonance resolved. The spectra are shifted vertically for clarity.

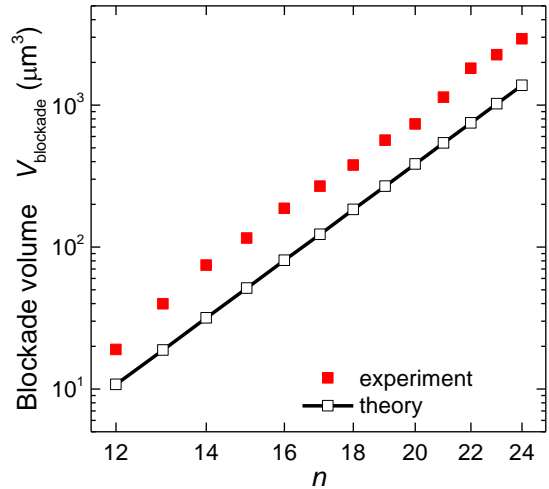

**Figure S4. Estimation of the blockade volume.** Squares give experimental data for excluded volume $V_{blockade}$ calculated from the $S_n$ values, solid line represents results of theoretical calculation. For $n=24$ the corresponding critical radius $R_c$ of the blockade volume amounts to about 10 μm.

| n | Fitted exciton energies $E_n$ (eV) | Positions of the peaks in the absorption spectrum (eV) |
|---|---|---|
| 2 | 2.1484 | 2.1472 |
| 3 | 2.16135 | 2.16120 |
| 4 | 2.16609 | 2.16604 |
| 5 | 2.16829 | 2.16827 |
| 6 | 2.16948 | 2.16946 |
| 7 | 2.170182 | 2.170172 |
| 8 | 2.170635 | 2.170628 |
| 9 | 2.170944 | 2.170939 |
| 10 | 2.171163 | 2.171159 |
| 11 | 2.171324 | 2.171322 |
| 12 | 2.171446 | 2.171444 |
| 13 | 2.1715410 | 2.171539 |
| 14 | 2.1716159 | 2.171614 |
| 15 | 2.1716758 | 2.171674 |
| 16 | 2.1717248 | 2.171723 |
| 17 | 2.1717653 | 2.171764 |
| 18 | 2.1717989 | 2.1717990 |
| 19 | 2.1718270 | 2.1718275 |
| 20 | 2.1718515 | 2.1718519 |
| 21 | 2.1718724 | 2.1718730 |
| 22 | 2.1718906 | 2.1718909 |
| 23 | 2.1719068 | 2.1719070 |
| 24 | 2.1719202 | 2.1719209 |
| 25 | 2.1719335 | 2.1719337 |

**Table S1.** Right column: Experimental positions of the resonances of P excitons in $Cu_2O$ from a 34μm sample at 1.2K. Left column: corrected values $E_n$ are calculated by unfolding the phonon background from the measured absorption spectrum according to Eq. S1.